\documentclass[conference,10pt]{IEEEtran}
\IEEEoverridecommandlockouts
\usepackage{cite}
\usepackage{amsmath,amssymb,amsfonts}
\usepackage{algorithmic}
\usepackage{graphicx}
\usepackage{textcomp}
\usepackage{xcolor}
\usepackage{array}
\usepackage{multirow}
\usepackage{enumitem}
\usepackage{makecell}

\newcolumntype{L}{>{\raggedright\arraybackslash}m{3.7 cm}}
\newcolumntype{V}{>{\raggedright\arraybackslash}m{2.5 cm}}
\newcolumntype{B}{>{\raggedright\arraybackslash}m{2 cm}}
\newcolumntype{e}{>{\centering\arraybackslash}m{0.5 cm}}
\newcolumntype{f}{>{\centering\arraybackslash}m{0.8 cm}}
\newcolumntype{h}{>{\centering\arraybackslash}m{0.2 cm}}
\newcolumntype{K}{>{\raggedright\arraybackslash}m{1.1 cm}}
\newcolumntype{k}{>{\raggedright\arraybackslash}m{1.5 cm}}
\newcolumntype{g}{>{\raggedright\arraybackslash}m{1.3 cm}}

\def\BibTeX{{\rm B\kern-.05em{\sc i\kern-.025em b}\kern-.08em
    T\kern-.1667em\lower.7ex\hbox{E}\kern-.125emX}}
\usepackage[letterpaper, portrait, margin=0.7in, bmargin=0.95in]{geometry}

\usepackage{fancyhdr}
\fancypagestyle{firststyle}{\fancyhf{}
	\fancyhead[L]{Dipankar Shakya, Hitesh Poddar, and Theodore S. Rappaport, ``Sub-Terahertz Sliding Correlator Channel Sounder with Absolute Timing using Precision Time Protocol over Wi-Fi",\textit{ in GLOBECOM 2023 - 2023 IEEE Global Communications Conference, }Kuala Lumpur, Malaysia, Dec. 2023, pp. 1--6.}
}

\begin{document}
\bstctlcite{IEEEtran:BSTcontrol}
\title{A Sub-Terahertz Sliding Correlator Channel Sounder with Absolute Timing using Precision Time Protocol over Wi-Fi}

\author{\IEEEauthorblockN{Dipankar Shakya, Hitesh Poddar, and Theodore S. Rappaport}
	\IEEEauthorblockA{\textit{NYU WIRELESS, Tandon School of Engineering, New York University, USA}\\
		\{dshakya, hiteshp, tsr\}@nyu.edu}
\thanks{This research is supported by the New York University (NYU) WIRELESS Industrial Affiliates Program, a grant from the NYU Research Catalyst Prize, and National Science Foundation (NSF) Research Grants: 2216332.}
}

\maketitle

\linespread{1.05}

\thispagestyle{firststyle}

\begin{abstract}
     Radio channels at mmWave and sub-THz frequencies for 5G and 6G communications offer large channel bandwidths (hundreds of MHz to several GHz) to achieve multi-Gbps data rates. Accurate modeling of the radio channel for these wide bandwidths requires capturing the absolute timing of multipath component (MPC) propagation delays with sub-nanosecond accuracy. Achieving such timing accuracy is challenging due to clock drift in untethered transmitter (TX) and receiver (RX) clocks used in time-domain channel sounders, yet will become vital in many future 6G applications. This paper proposes a novel solution utilizing precision time protocol (PTP) and periodic drift correction to achieve absolute timing for MPCs in power delay profiles (PDPs) --captured as discrete samples using sliding correlation channel sounders. Two RaspberryPi computers are programmed to implement PTP over a dedicated Wi-Fi link and synchronize the TX and RX Rubidium clocks continuously every second. This synchronization minimizes clock drift, reducing PDP sample drift to 150 samples/hour, compared to several thousand samples/hour without synchronization. Additionally, a periodic drift correction algorithm is applied to eliminate PDP sample drift and achieve sub-nanosecond timing accuracy for MPC delays. The achieved synchronicity eliminates the need for tedious and sometimes inaccurate ray tracing to synthesize omnidirectional PDPs from directional measurements. The presented solution shows promise in myriad applications, including precise position location and distributed systems that require sub-nanosecond timing accuracy and synchronization among components.
\end{abstract}



\section{Introduction}
The empirical characterization of mmWave and sub-THz wireless channels in temporal and spatial domains is vital to simulate and design future wireless networks with vast bandwidths and multi-Gbps data rates\cite{rappaport2022cup}. Many futuristic applications, such as precise position location, wireless cognition, and sensing, require sub-nanosecond accuracy in both measurements and models for characterizing the spatio-temporal behavior of wireless channels\cite{akyildiz2022tc,Kanhere2023icc}. To make such accurate measurements and models with high-resolution of MPC delays, it is vital to use synchronized, untethered channel sounders capable of maintaining sub-nanosecond accuracy with standard time references\cite{rappaport2022cup,novotny2016pamta,Mac2017sac}.

Time-domain channel sounders with untethered TX and RX typically rely on atomic standards, such as Cesium (Cs) or Rubidium (Rb) for synchronized measurement of the channel impulse response. However, differences in the phase/frequency output of the asynchronous system components connected to the two independent atomic clocks can cause a drift in the time keeping between TX and RX systems, which is termed as \textit{time drift}. 
This can lead to undersired offsets in MPC delays of recorded PDPs that accumulate over time \cite{novotny2016pamta,Mac2017sac}. The MPC delay offsets in time periodic channel sounders appear as a circular shift in the recorded PDP samples thereby causing an error in the MPCs' absolute propagation delay. Furthermore, time-domain channel sounders, such as one presented in \cite{Mac2017sac}, employ narrow half-power beamwidth (HPBW) horn antennas to measure directional PDPs in non-overlapping spatial steps within a 360$^{\circ}$ azimuth sweep. The accumulated offset in MPC delays within directional PDPs at different pointing angles over time can cause false or missed MPCs when omnidirectional PDPs are synthesized\cite{sun2015gc,Samimi2016tmtt}. As these omnidirectional PDPs are used by industry standard bodies to generate statistical models, the time accuracy of captured MPCs is crucial, and dependent on the TX and RX system clock synchronization\cite{rappaport2022cup}.       

The drift between TX and RX clocks in untethered time domain channel sounders is highlighted as a persistent problem in the literature\cite{novotny2016pamta,Mac2017sac,Kast2021tim,riley2008nist,Fatih2020ojap}. Authors in \cite{Kast2021tim} quantify the \textit{time drift} between TX and RX with deterministic and stochastic components ((3) in \cite{Kast2021tim}). Researchers in \cite{Fatih2020ojap} also observe significant \textit{time drift} and report ``ghost-multipath" detection or missed MPCs due to drift. Authors in \cite{novotny2016pamta,Mac2017sac} rely on periodic re-synchronization between TX and RX Rb clocks to minimize PDP shifting due to \textit{time drift}. GPS-based clock disciplining has also been widely used for distributed systems and outdoor applications such as UAVs or power grids\cite{qu2022spr}, but are unreliable under obstructed sky. Researchers also explore leveraging the network stack for internet or local area network (LAN)-based clock synchronization\cite{Mills1991tc,Aslam2022tii,chen2021usenix,Shakkottai2003icm}. VNA-based frequency-domain channel sounders avoid \textit{time drift} using co-located phase-synchronized source and receiver. However, such systems face challenges for long-range measurements with obstructions in between \cite{Mac2017sac}. Ray-tracing is an accurate but computationally complex method that synthesizes propagation paths and exact delays of MPCs using 3D models of the environment\cite{Kanhere2023icc,Lecci2020ita,Samimi2016tmtt,Ju2021jsac}. Researchers also record the rate of PDP shifting at different intervals during field measurements and estimate the \textit{time drift} between observations to time-align PDPs in post-processing\cite{ju2023icc,Fatih2020ojap,Mac2017sac}. These methods involve extensive post-processing and manual labor, which may induce human errors in processing the measured data. Thus, a reliable and automatic method to remove the \textit{time drift} will be a vast improvement.  

This paper details novel enhancements to the NYU WIRELESS channel sounder to achieve PDP time-alignment for directional PDPs. Key enhancements and observations are:
\begin{itemize}
	\item A RaspberryPi single-board computer added at the TX and RX implements IEEE 1588-2008 precision time protocol (PTP) \cite{ptp2020ieee} over Wi-Fi\cite{Aslam2022tii,Shakkottai2003icm}. (Section \ref{sxn:PTPsol})
	\item PTP reduces PDP sample drift to within 150 samples/hour from several thousand. This corresponds to a few samples in a 3-minute measurement sweep making absolute timing achievable. (Section \ref{sxn:Bgnd} and \ref{sxn:AT})
	\item Minimizing number of hops in a PTP network improves synchronization accuracy (Table \ref{tab:ptp})
	\item Periodic and automated PDP drift correction between measurement sweeps achieves absolute timing in channel sounding. (Section \ref{sxn:AT})
	\item Systematic implementation of the PDP drift correction and TX-RX (T-R) location change improves existing channel measurement procedures\cite{Mac2017sac,ju2023icc}. (Section \ref{sxn:AT})    
\end{itemize}
The method given here requires minimal additional hardware, making implementation simple and cost effective. This method solves the \textit{time drift} problem observed in previous channel sounder systems and offers promise for a method that can be used in other applications of mmWave and sub-THz measurement beyond channel sounding, such as distributed systems, radar, and precise position location.


\section{142 GHz Sliding Correlation Channel Sounder}\label{sxn:Bgnd}
\subsection{Sliding Correlation based Channel Sounding}
Sliding correlation based channel sounding uses cross-correlation between pseudo-random noise (PN) sequences to study radio propagation behavior in the time domain \cite{tsr2002ph,Shakya2021tcas2}. The TX uses binary phase shift keying to modulate an RF carrier with the PN sequence generated at a fast chip rate of $\alpha=$ 500 Mcps in a superheterodyne architecture \cite{Shakya2021tcas2}. At the RX, multiple replicas of the transmitted PN sequence are received at the antenna due to multipath and downconverted to the baseband as in-phase and quadrature components. Here, an identical PN sequence is generated at a slightly slower chip rate of $\beta=$ 499.9375 Mcps and is correlated with the downconverted received signal. The slight difference in the chip rates at the TX and RX causes the received signal to slide over the slower replica PN sequence to generate a time-dilated PDP with peaks where a PN auto-correlation is achieved. 

\begin{figure}[htbp]
	\centering
	\includegraphics[width=0.48\textwidth]{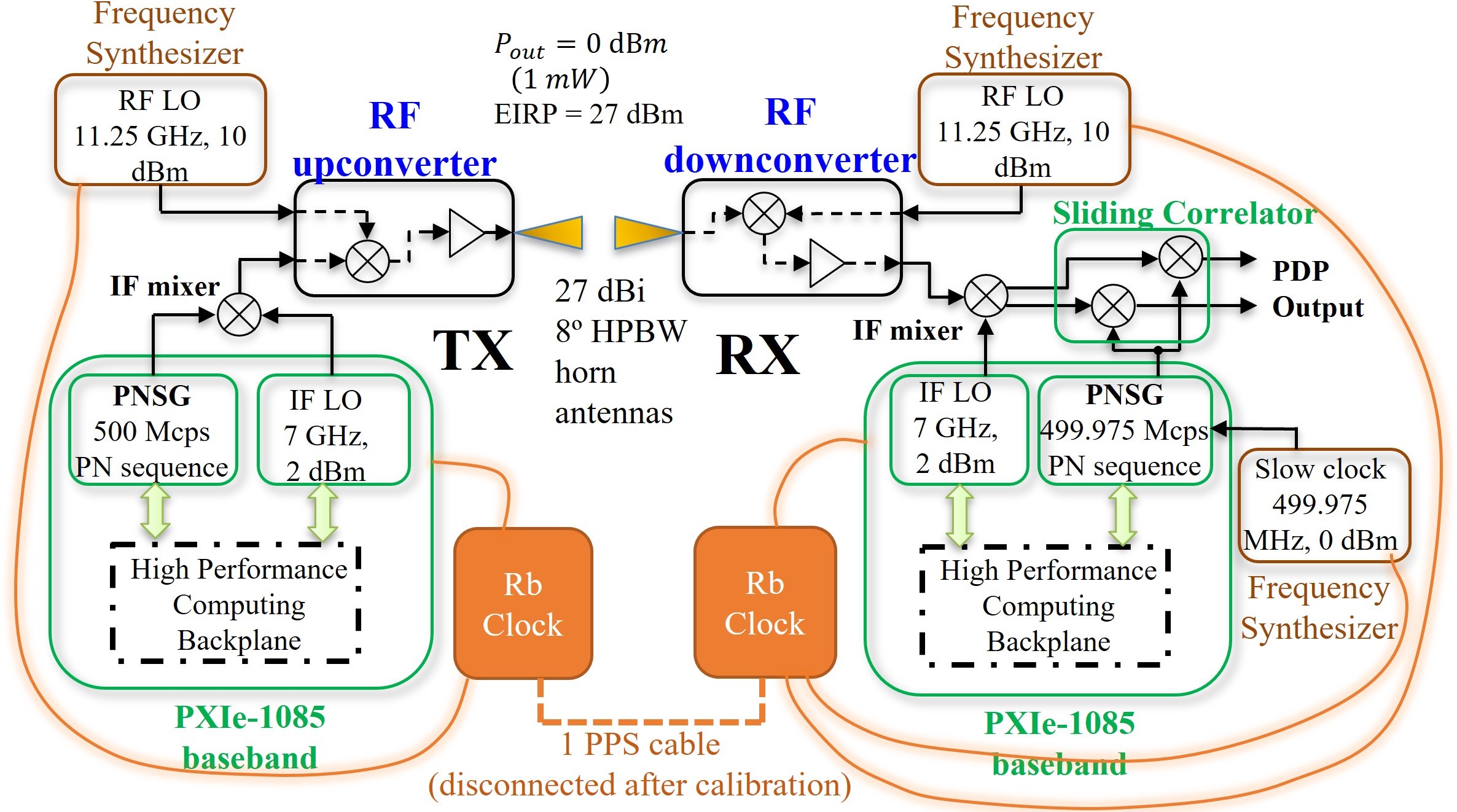}
	\vspace{-2 em}\caption{System block diagram of 142 GHz channel sounder. At the TX, a 500 Mcps PN sequence is upconverted and transmitted; At the RX, sliding correlation is performed using an identical PN sequence at 499.95 Mcps\cite{Mac2017sac}. The TX and RX are each synchronized with Rubidium clocks.}
	\label{fig:Sys_blk}
	\vspace{-1 em}
\end{figure}

The sliding correlation causes a time-dilation of the PDP by a slide factor, $\gamma$, resulting in a processing gain ($G_p$), as in \eqref{eq:pgain}, and bandwidth compression while rejecting external interference\cite{tsr2002ph,Shakya2021tcas2}. Moreover, time-dilation eases nanosecond clock synchronization requirements allowing PTP over Wi-Fi to synchronize untethered TX and RX. Fig.\ref{fig:Sys_blk} shows the block diagram of the 142 GHz sliding correlation channel sounder at NYU.

\begin{equation}
	\vspace{-1 em}
\label{eq:pgain}
\begin{split}
G_{p} \text{ [dB]}&= 10\log_{10}(\gamma) \text{ [dB]},\\
\gamma&= \alpha/(\alpha-\beta) ,
\end{split}
\vspace{-1 em}
\end{equation} 
 

\subsection{Rubidium Clocks}
Atomic clock standards, such as the Rubidium (Rb) or Cesium (Cs), can provide stable clock outputs for use between two physically separated systems. The Rb standard trains a crystal oscillator to the fine transition of electrons in Rb-87 atoms, implemented as a rubidium lamp and detected by photodetectors \cite{stanfordrs2015srs}; This standard is inexpensive and widely used owing to a high accuracy of typically $5\times10^{-3}$ parts-per-million (ppm) after $\sim$ 10 years of operation. 
Commercial Rb clocks, such as the FS725 developed by Stanford Research Systems can be synchronized using the `1-pulse-per-second' (1 PPS) signal. The 1 PPS is a beat signal that is comprised of a 10-20 $\mu$s pulse with sharp rising and falling edges repeating once every second. The 1 PPS-Out port on a `master' Rb clock is typically cable-connected to the 1 PPS-In port on a `slave' Rb clock. The slave clock adjusts its phase and frequency to synchronize with the 1 PPS signal master. 

Besides the 1 PPS signal, the Rb clocks output a 10 MHz reference (REF) signal that asynchronous components of the channel sounder synchronize to. At the TX (or RX) baseband in Fig.\ref{fig:Sys_blk}, the signal generator for the $\alpha$ (or $\beta$) rate PN sequence and the IF LOs synchronize to the local TX (or RX) Rb clock via the 10 MHz REF input to the PXIe-1085. The RF LO frequency synthesizers are also synchronized to the Rb clock with a 10 MHz REF connection. Additionally, at the RX baseband, the slow clock frequency synthesizer synchronizes to the RX Rb clock via a 10 MHz REF input. 

\subsection{Clock Accuracy} 
Manufacturers specify an accuracy for clocks in parts-per-million (ppm). An accuracy specification of $\pm$ 25 ppm (e.g., PXIe-1085 system clock\cite{pxie2015ni}) implies that for every one million oscillations at a specified frequency, the clock can have an error of $\pm$ 25 oscillations. For a 10 MHz clock, the output frequency can vary between 10 MHz $\pm$ 250 Hz with a 25 ppm accuracy specification. The $5\times10^{-3}$ ppm \cite{stanfordrs2015srs} clock accuracy on the FS725 Rb clock results in synchronized local operation at the TX/RX. However, wireless synchronization between Rb clocks to prevent PDPs from drifting is a challenge. We show how to overcome this challenge in Sections \ref{sxn:PTPsol} and \ref{sxn:AT}.  



\subsection{PDP capture and recording}\label{sec:PDPcapt}
Time-dilated PDPs resulting from sliding correlation at the RX are captured via the oscilloscope. At each unique antenna pointing direction, 20 PDPs are collected and averaged to generate one directional PDP, which is stored with a log of the system state and TX/RX pointing angles. Each PDP captures the radio channel over the time-dilated duration of the baseband PN sequence length at the RX oscilloscope sampling rate (one PDP = 81880 samples = 2047 [chips]$\times$2e-9 s [chip width]$\times$8000 [$\gamma$]$\times$2.5e6 sps [sampling rate] $\equiv$ 32.752 ms)\cite{Mac2017sac}. The TX and RX sweep the azimuth in antenna HPBW steps at different elevation angles using rotatable gimbals. In the field, a $360 ^{\circ}$ TX/RX azimuth sweep corresponds to a measurement sweep and lasts $\sim$3 mins \cite{Mac2017sac, ju2023icc}.     

\section{Impact of Time Drift in the Sliding Correlation Channel Sounder}\label{sxn:tdrift}
\begin{figure}[htbp]
	\centering
	\includegraphics[width=0.45\textwidth]{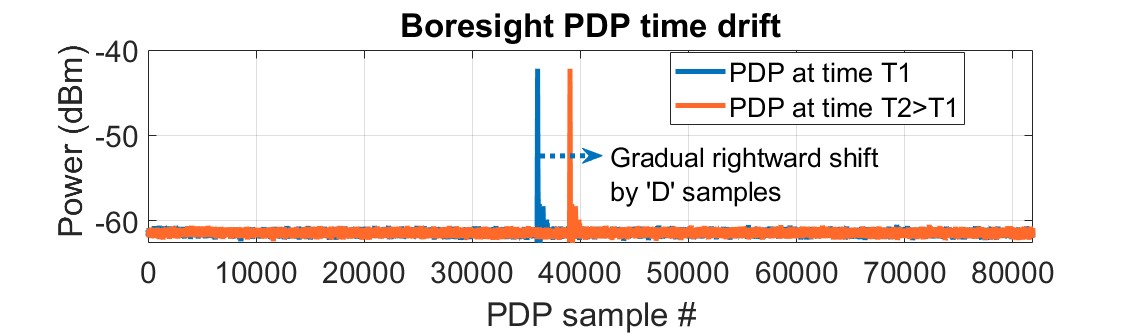}
	\vspace{-1 em}\caption{Example of MPC delay offset due time drift on a boresight PDP containing a single MPC at an arbitrary sample \# 36000. With time the boresight peak is illustrated to shift rightward and appear at sample \# 39000. Here, $D= 3000$. \newline}
	\label{fig:EgPDPdrift}
	\vspace{-2.5 em}
\end{figure}

In sliding correlation channel sounders the impact of \textit{time drift} is observed as the gradual circular shifting of the recorded samples of the time-dilated PDP. Slight differences between the performances of TX and RX system Rb clocks, LO, and internal baseband processor clocks cause the \textit{time drift} \cite{Mac2017sac,Kast2021tim,Fatih2020ojap}. For example, assuming a TX and RX at boresight with a single line-of-sight (LOS) propagation path between antennas, the time-dilated PDP captured at the RX with an oscilloscope contains a single peak corresponding to the LOS MPC. The RX oscilloscope captures the same PDP periodically at 2.5 Msps collecting 81880 samples when a trigger occurs every 32.752 ms, assuming the channel is static. 32.752 ms is the slide time to capture one full PDP spanning the 4094 ns PN sequence duration (2047 [chips]$\times$ 2 ns [chip width], Section \ref{sec:PDPcapt}). The PDP comprised of the 81880 samples would appear as shown in Fig. \ref{fig:EgPDPdrift}, with the peak occurring at, say, sample \#36000 as an arbitrary example. Over time, the MPC peak is observed to shift right by D=3000 samples, which causes an error in the actual MPC propagation delay by 150 ns (one PDP sample = 4094 ns / 81880 samples). 

\begin{figure}[htbp]
	\centering
	\includegraphics[width=0.48\textwidth]{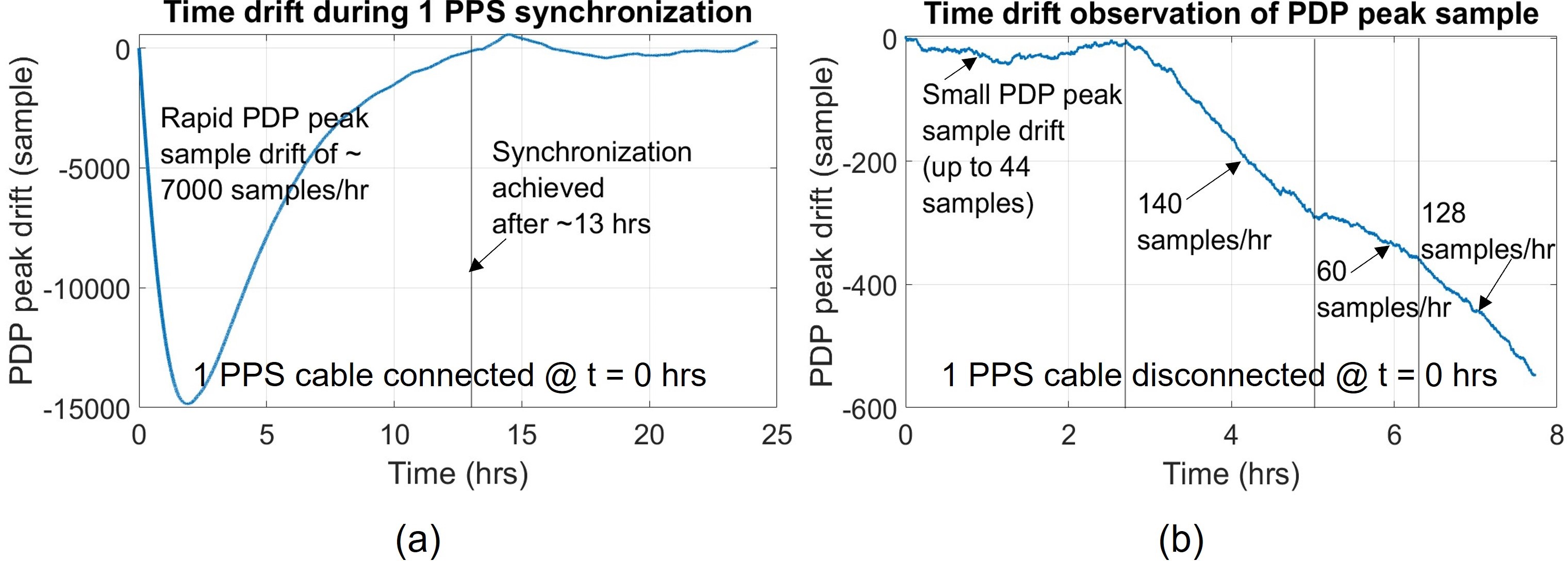}
	\vspace{-2 em}\caption{Observation of the time drift: (a) during 1 PPS synchronization when the MPC peak drifts at a rate of $\sim$7000 samples/hr. (b) after synchronization when the PPS cable is disconnected.}
	\label{fig:sync_drift}
	\vspace{-1 em}
\end{figure}

The FS725 Rb clocks typically indicate a 1 PPS lock after 15-20 minutes of cable-connected master-slave training. This lock indicates that the slave Rb clock has adjusted its frequency and phase to match the master Rb clock 1 PPS signal to within 1 $\mu$s. In reality, it requires 10-13 hours of cable-connected Rb clock training, in agreement with \cite{Kast2021tim}, to recover from a fast PDP \textit{time drift} of $\sim$7000 samples/hr, as shown in Fig. \ref{fig:sync_drift} (a), which may occur after a few hours of operation without the 1 PPS cable. Observations of an identical PDP --corresponding to a T-R antenna pointing direction --at different time instants allow the automatic capturing of the PDP \textit{time drift}. A downward transition with time in Fig. \ref{fig:sync_drift} would represent a drift of the MPC peak(s) to the left and vice-versa in Fig. \ref{fig:EgPDPdrift}.

Although a 1 PPS signal is used to bring the two system clocks in sync, the PDPs exhibit a fast drift typically 2-3 hours after disconnecting the PPS cable causing the drift to transition from 44 to 140 samples/hr, as observed in Fig.\ref{fig:sync_drift} (b). The \textit{time drift rate} varies with time due to different inaccuracies of clock driven components of the channel sounder. The drift rates, however, remain relatively linear between rate transitions, as indicated in the different zones in Fig. \ref{fig:sync_drift} (b). In channel measurement campaigns with untethered TX and RX that span several days and many hours/day, the accumulated drift can cause misinterpretations in the delays of MPCs and their powers captured by PDPs collected throughout a day. 

\renewcommand{\arraystretch}{1.2}
\begin{table*}[htbp]
	\centering
	\caption{PTP configurations tested}
	\vspace{-1 em}
	\begin{tabular}{|h|V|e|K|L|K|L|k|}
		\hline
		\textbf{\#} & \textbf{Configuration} & \textbf{Hops} & \textbf{PTP type} & \textbf{Description} & \textbf{Sync.\newline Accuracy} & \textbf{Notes} & \textbf{Linux libraries} \\
		\hline
		1     & PTP over two bridged Wi-Fi routers & 3 & hardware & \vspace{-2\baselineskip} \noindent{\begin{itemize}[leftmargin=0.15cm]
				\item RPis connected to different WiFi routers via ethernet \cite{Geerling2022blog}.
				\item WiFi routers bridged into common WLAN \cite{Clark2023git}.
				\vspace*{-2\baselineskip}
		\end{itemize}}  & 30-40 ms & \vspace{-1\baselineskip}
		\noindent{\begin{itemize}[leftmargin=0.15cm]
				\item Wi-Fi bridging masks RPi MAC and device name.
				\item Use static IPs and device MAC when configuring router.
				\vspace*{-2\baselineskip}
		\end{itemize}} & \texttt{linuxptp} \\
		\hline
		2     & PTP : \newline{}RPi master on ethernet; \newline{}RPi slave on WLAN & 2     & hardware-software hybrid & \vspace{-1\baselineskip}
		\noindent{\begin{itemize}[leftmargin=0.15cm]
				\item RPi master conneted to LAN port of Wi-Fi router for hardware timestamping \cite{Clark2023git}.
				\item RPi slave connected to WLAN from Wi-Fi router.
				\vspace*{-\baselineskip}
		\end{itemize}} & $\sim$50 $\mu$s & \vspace{-1\baselineskip}
		\noindent{\begin{itemize}[leftmargin=0.15cm]
				\item Turn OFF Wi-Fi power management on RPi slave to improve accuracy\cite{Henderson2023blog}.
				\item RPi slave ethernet port dummy-terminated*.
				\vspace*{-\baselineskip}
		\end{itemize}} & \multirow{3}{2cm}{\newline{}\texttt{linuxptp},\newline{}\texttt{ptpd}}\\
		\cline{1-7} 3     & PTP with RPis on common WLAN & 2     & \multirow{2}{1cm}{\\software} & \vspace{-1\baselineskip}
		\noindent{\begin{itemize}[leftmargin=0.15cm]
				\item  RPi master and slave connect to WLAN from one router
				\vspace*{-\baselineskip}
		\end{itemize}} & $\sim$ 80 $\mu$s & \multirow{2}{3cm}{\vspace*{-1.2\baselineskip}
			\noindent{\begin{itemize}[leftmargin=0.15cm]
					\item Turn OFF Wi-Fi power management on RPis to improve accuracy.
					\item Both RPi ethernet ports dummy-terminated*.
					\vspace*{-\baselineskip}
		\end{itemize}}} & \\
		\cline{1-3} \cline{5-6} 4     & PTP: \newline RPi master hotspot; \newline RPi slave on hotspot & 1     & &\vspace{-1\baselineskip}
		\noindent{\begin{itemize}[leftmargin=0.15cm]
				\item RPi master creates a WLAN hotspot network.
				\item RPi slave on created hotspot.
				\vspace*{-\baselineskip}
		\end{itemize}} & $\sim$ 5 $\mu$s &  & \\ 
		\hline
		\multicolumn{8}{c}{*dummy-terminated ethernet connected to internet/wall port on router. Termination required for NIC PPS output.}
	\end{tabular}%
	\vspace*{-1.5\baselineskip}
	\label{tab:ptp}%
\end{table*}%
\renewcommand{\arraystretch}{1}

\section{The Precision Time Protocol Implementation}\label{sxn:PTPsol}
\subsection{Overview}
\begin{figure}[htbp]
	\centering
	\includegraphics[width=0.48\textwidth]{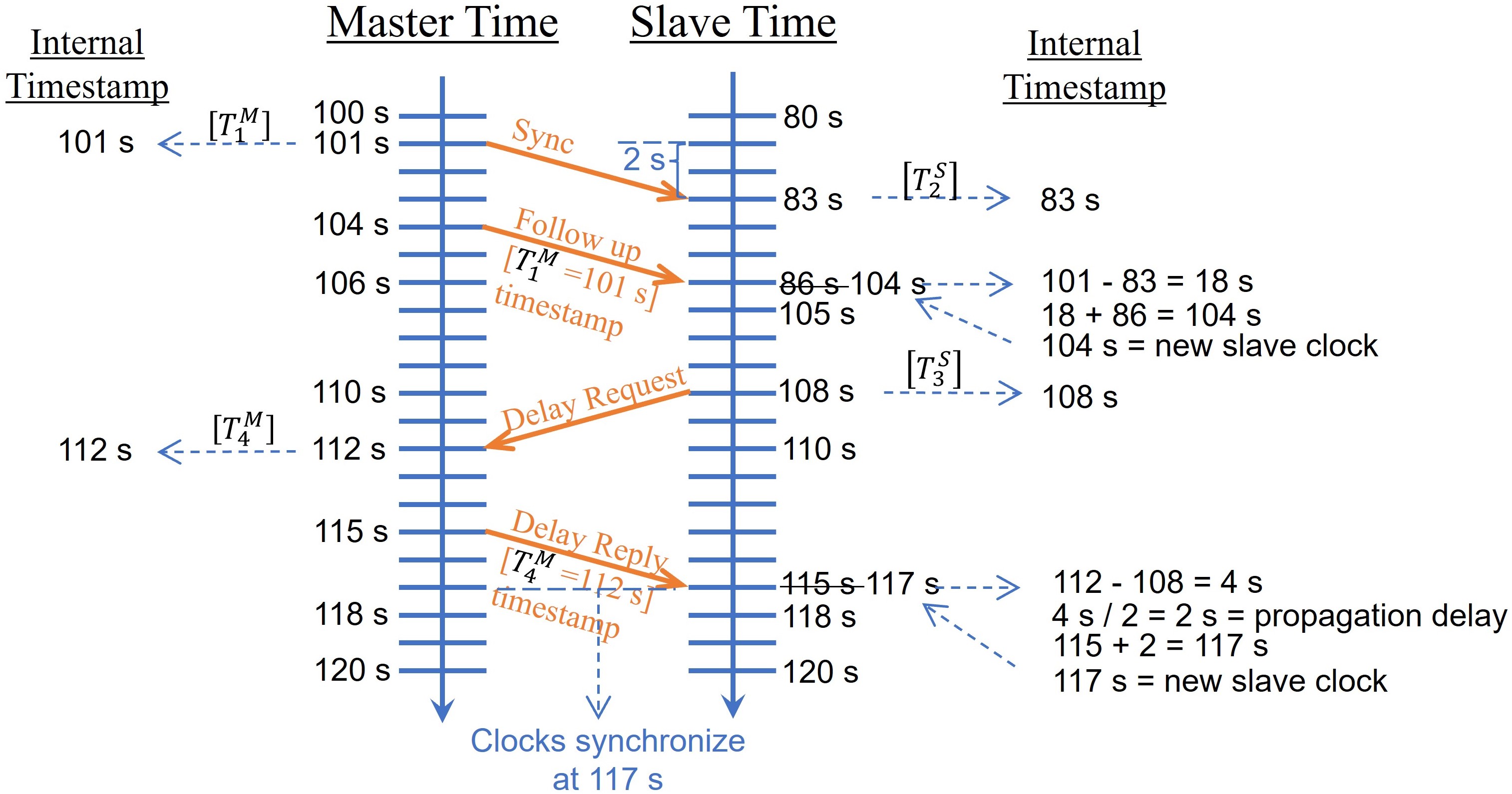}
	\vspace{-2 em}\caption{Example showcasing the fundamental synchronization algorithm employed in PTP that uses four timestamp packet transactions to determine clock offset (20 s), including propagation delay (2 s).}
	\label{fig:PTPop}
	\vspace{-1 em}
\end{figure}

Precision time protocol (PTP) is primarily used in wired local computer networks for synchronization through nanosecond resolution time-stamping of PTP packets\cite{chen2021usenix}. In contrast to network time protocol (NTP)\cite{Mills1991tc} that synchronizes to an NTP server with milliseconds accuracy over internet, PTP uses hardware timestamps from an on-board network interface chip (NIC) that is communicated to other devices in a LAN. In Fig. \ref{fig:PTPop} an example helps illuminate PTP time synchronization for a single `master' and `slave'. Assume a 1 s resolution time stamping with an arbitrary 2 s signal propagation delay between the nodes. In reality, these time values would be on the order of ns or $\mu$s and arbitrary values are used to simplify the example \cite{chen2021usenix}. For the master's clock, assume 100 s have elapsed. Likewise, the unsynchronized slave might have a clock with 80s elapsed (i.e. 20 s offset from master). The master in Fig. \ref{fig:PTPop} initiates the PTP synchronization by recording a timestamp ($T_1^M = 101 s$) while sending out a `Sync' packet. Upon receiving it 2 s later, the slave also records a timestamp ($T_2^S = 83 s$) and waits for a `Follow-up' packet. After clock cycles, master transmits $T_1^M$ the Follow-up packet; When received, the slave calculates $T_1^M-T_2^S$ and adds it to its current time. This partial correction adjusts the slave clock from $86 s$ to $104 s$ in Fig. \ref{fig:PTPop}, but is missing a propagation delay component (here, 2 s). Hence, the slave records $T_3^S = 108 s$ timestamp and sends out a `Delay-request'. The master records $T_4^M = 112 s$ and sends back a `Delay-reply' with $T_4^M$. Assuming a reciprocal channel where Delay-request and Delay-reply propagate in equal duration, the propagation delay is calculated as $(T_4^M-T_3^S)/2$ and added to the slave time to achieve synchronization at 117 s.

\subsection{PTP on the sliding-correlation channel sounder}
The RaspberryPi (RPi) compute module 4 (CM4) has a built-in NIC capable of PTP hardware time-stamping at the physical layer of the network stack. While dedicated PTP devices and grand-master clocks are commercially available for large networks \cite{chen2021usenix,Aslam2022tii}, the RPi offers a compact and effective PTP solution when used with Rb clocks\cite{Geerling2022blog}. 
The RPi CM4 NIC can synchronize its system clock to a 1 PPS signal or output it. However, as hardware PTP is primarily designed for wired networks, the NIC is only connected to the ethernet port on the RPi CM4. Nonetheless, Rb clocks on the channel sounder require wireless clock synchronization for real-world radio propagation measurements\cite{Samimi2016tmtt,Ju2021jsac}. Some workarounds using bridged Wi-Fi routers\cite{Kulau2016dcoss} or software-based time-stamping leveraging layers above the physical-layer are tested (Table \ref{tab:ptp}). `Sync. Accuracy' in Table \ref{tab:ptp} represents the difference between RPi master and slave timestamps, as reported by the RPi slave: $T_1^M-T_2^S$ in Fig. \ref{fig:PTPop}. 

\begin{figure}[htbp]
	\centering
	\includegraphics[width=0.48\textwidth]{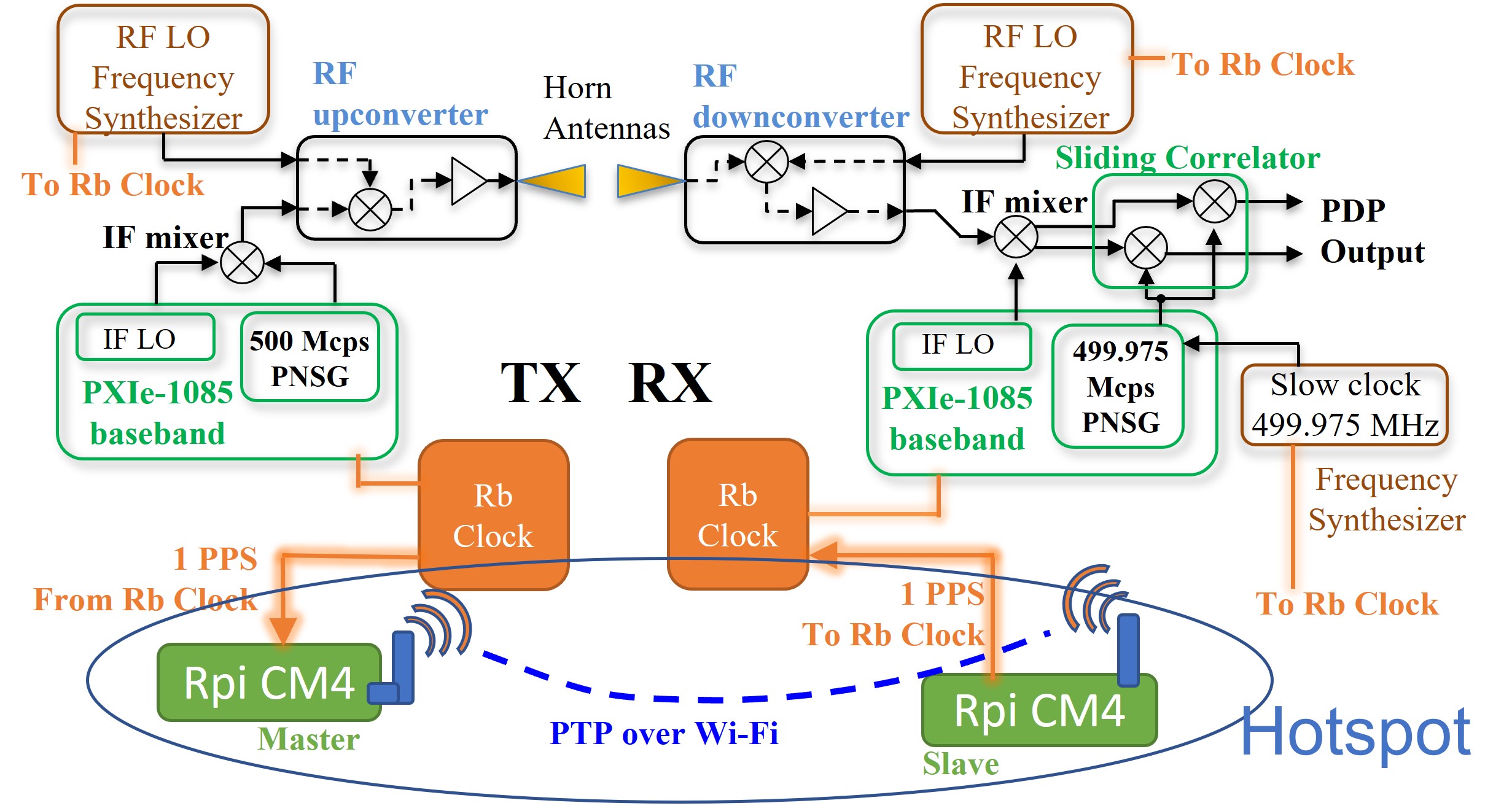}
	\vspace{-2 em}\caption{The Rb Clock at the channel sounder TX synchronizes a RaspberryPi that implements PTP hardware time-stamping via its NIC. \newline}
	\label{fig:142ChSdr_AT}
	\vspace{-1 em}
\end{figure}

As illustrated in Fig. \ref{fig:142ChSdr_AT}, two RPis establish a wireless medium for clock synchronization using PTP. The RPi master synchronizes to the 1 PPS signal from the TX Rb clock. The slave RPi clock is synchronized to the master RPi via PTP and generates a 1 PPS signal to synchronize the RX Rb clock. Motion and changes in the wireless medium can vary the synchronization accuracy between master and slave RPi clocks, compared to a cable connection, which is accurate to a sub-nanosecond\cite{Geerling2022blog}. A software-based PTP is implemented over a dedicated Wi-Fi hotspot link created by the RPi master at the TX. The RPi slave at the RX connects to this hotspot to maintain a single hop for greater accuracy, based on the results in Table \ref{tab:ptp}. The RPi master and slave use their on-board Wi-Fi modules to establish and maintain the link. 

\begin{figure}[htbp]
	\centering
	\includegraphics[width=0.48\textwidth]{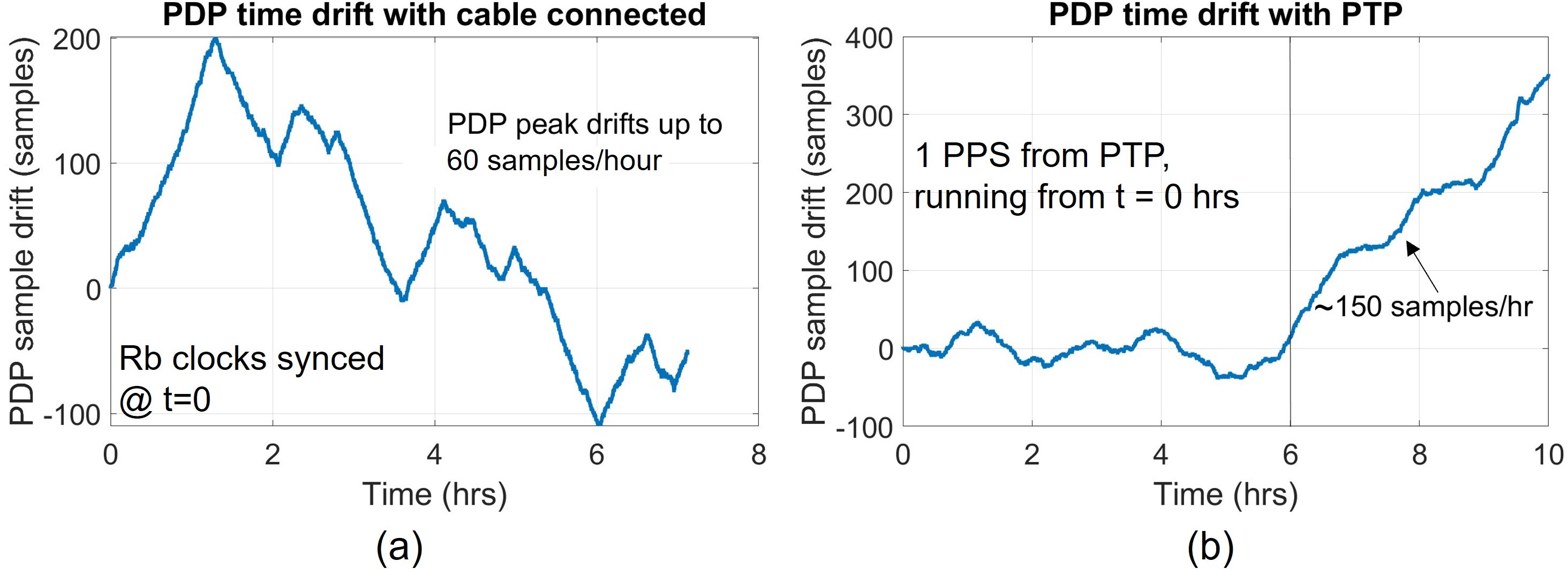}
	\vspace{-2 em}\caption{Observation of PDP \textit{time drift}: (a) 1 PPS cable connecting Rb clocks (b) PTP over Wi-Fi implemented (cable disconnected), as in Fig. \ref{fig:142ChSdr_AT}}
	\label{fig:Cab_PTP}
	\vspace{-1 em}
\end{figure}
    
Fig. \ref{fig:Cab_PTP} shows the PDP drift comparison when a 1 PPS cable is directly connected between Rb clocks and PTP is implemented over dedicated Wi-Fi. Even with the cable connected, PDP \textit{time drift} is observed up to 60 samples/hour. The PTP synchronized Rb clocks showed almost no PDP drift for up to six hours, as seen in Fig. \ref{fig:Cab_PTP}(b). Then, the system transitioned to a PDP \textit{time drift} up to 150 samples/hour due to change in the synhronization accuracy of the PTP link from motion in the wireless medium. Though inferior to cable-connected operation, the effect of \textit{time drift} on PDPs is significantly minimized using PTP as compared to drifts in Fig. \ref{fig:sync_drift}. Hence, additional processing (Section \ref{sxn:AT}) is required to achieve absolute timing accuracy.

\section{Periodic PDP drift correction for absolute timing}\label{sxn:AT}
\begin{figure}[htbp]
	\centering
	\includegraphics[width=0.48\textwidth]{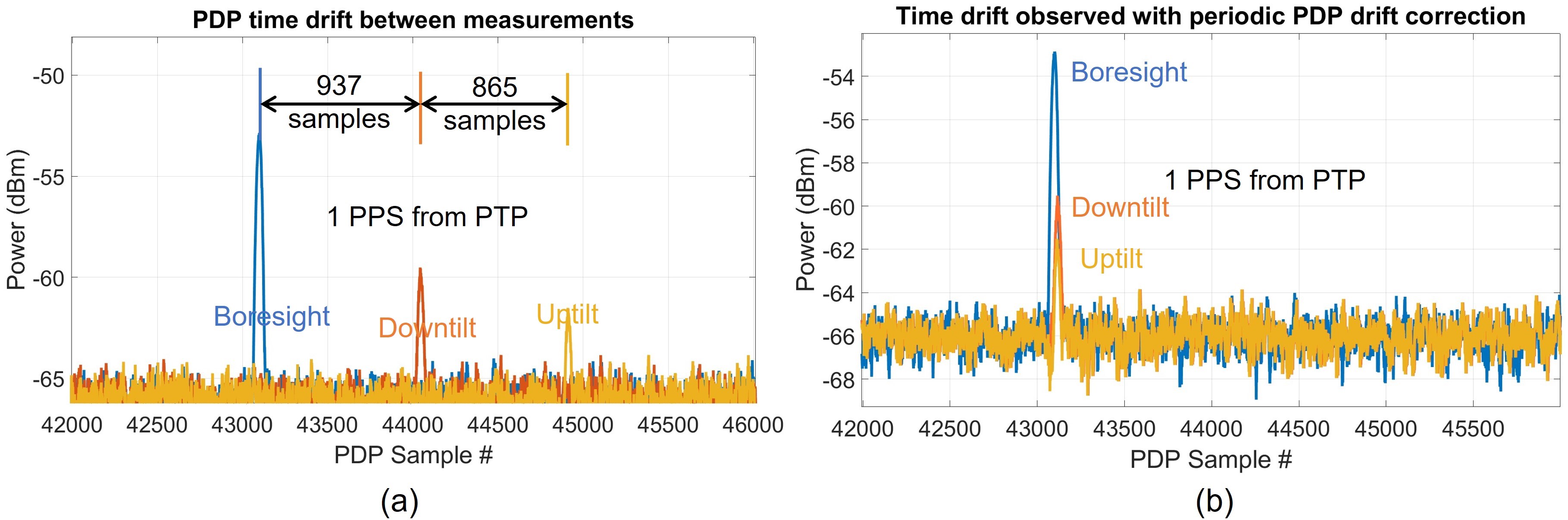}
	\vspace{-2 em}\caption{PDP \textit{time drift} between T-R boresight, RX 8$^\circ$ downtilt, and RX 8$^\circ$ uptilt measurement sweeps for: (a) Only PTP synchronization  (b) PTP synchronization with periodic PDP drift correction between measurements. \newline}
	\label{fig:PDP_SA}
	\vspace{-1 em}
\end{figure}

The PDP \textit{time drift} is observable between successive 3-minute measurement sweeps during radio propagation measurement campaigns without PTP implementation. It is most prominent when the TX and RX move to new locations in a campaign day (e.g. Fig. 1 in \cite{Ju2021jsac} shows the indoor office location map with T-R locations). Several measurement sweeps are performed at a T-R location pair before TX and/or RX are moved to a different spot. Fig. \ref{fig:PDP_SA} (a) shows the PDP peak, corresponding to the same LOS MPC, drift between successive measurement sweeps where the RX is at boresight, downtilted, then uptilted by the antenna HPBW (here, 8$^\circ$). Such large \textit{time drift} occurs when several hours have passed since the T-R Rb clocks were last fully synchronized.
 
A periodic PDP drift correction is implemented with PTP to re-align PDPs between measurement sweeps. Although a small drift of few tens of samples in an 81880 sample PDP may exist, it is considerable for statistical channel modeling (e.g. 18 samples in 3 minutes for a 360 samples/hour PDP \textit{time drift} is $\sim$1 ns of drift). Further, with PTP synchronization the drift within a measurement is only a few samples. Fig.\ref{fig:PDP_SA} (b) shows the PDP peaks realigned between measurement sweeps with periodic correction, thus, achieving sub-nanosecond accurate absolute timing for MPCs between measurements. Algorithm 1 details the periodic PDP drift correction. 

Time-drift tracking is leveraged to compensate \textit{time drift} that accumulates during a TX/RX location change within a measurement site, on the same day. \textit{Time drift} may be assumed linear within the span of a location change that lasts tens of minutes such that impact of external factors including temperature or motion of the channel sounder are nominal, based on observations in Fig. \ref{fig:sync_drift}  and \ref{fig:Cab_PTP}. Additional experiments, such as moving the cart on the lab floor and over bumpy cable tracks, showed no significant change in \textit{time drift rate}. Algorithm 2 presents steps for a systematic location change.

The periodic re-alignment and systematic location change algorithms prevent PDP \textit{time drift} from accumulating and causing significant modeling errors. The steps presented in Algorithm 1 and 2 improve the channel measurement, calibration, and drift tracking algorithms taught in \cite{Mac2017sac} and \cite{ju2022nyu}. The successful implementation of both algorithms rely on proper calibration completed at the start of each measurement day\cite{ju2022nyu,Mac2017sac,Samimi2016tmtt}. Moreover, these algorithms with PTP over Wi-Fi obviate the tedious post processing relying on ray-tracing or time drift tracking. The measurements can additionally stand as a reference for calibrating ray-tracers \cite{Kanhere2023icc}.   


\begin{table}[htbp]
	\centering
	\begin{tabular}{p{0.45\textwidth}}
		\hline
		\textbf{Algorithm 1: Periodic PDP drift correction} \\
		\hline
		\vspace{1\baselineskip} \vspace{-0.5 cm}
		\textit{Step 1}: Find the strongest MPC direction for a T-R location. \newline \hspace*{9mm}This MPC shall be referred as `reference MPC'.\\
		\textit{Step 2}: Record the MPC peak position in the PDP as $P1$.\\
		\textit{Step 3}: Record the TX and RX azimuths and elevations. \newline \hspace*{9mm}This position shall be referred as `reference position'.\\
		\textit{Step 4}: Begin a standard measurement sweep for the channel sounder.\\
		\textit{Step 5}: After the sweep completes, return TX and RX to \newline \hspace*{9mm}reference positions. \\
		\textit{Step 6}: Re-record MPC peak position for the reference MPC as $P2$. \\
		\textit{Step 7}: Circular shift the PDP by $P1 - P2$. \\
		\textit{Step 8}: Repeat from \textit{Step 4} until all measurement sweeps are complete. \\
		\hline
	\end{tabular}%
	\label{tab:alg1}%
	\vspace{-1 em}
\end{table}%

\begin{table}[htbp]
	\centering
	\begin{tabular}{p{0.45\textwidth}}
		\hline
		\textbf{Algorithm 2: Systematic Location Change} \\
		\hline
		\vspace{1\baselineskip} \vspace{-0.5 cm}
		\textit{Step 1}: Before starting re-location, initiate time-drift tracking\cite{ju2022nyu}. \newline \hspace*{9mm}Run for 15 minutes.\\
		\textit{Step 2}: Determine the PDP drift rate (samples/sec) and record as $Rate$.\\
		\textit{Step 3}: Record a timestamp $T1$ at the end of time-drift tracking.\\
		\textit{Step 4}: Move the TX and RX to their new locations. \newline \hspace*{9mm}Stabilize the systems and prepare for T-R measurement sweeps.\\
		\textit{Step 5}: Record a timestamp $T2$ before starting measurement sweep. \newline \hspace*{9mm}Calculate approximate samples shifted as: \newline \hspace*{9mm}$S = (T1-T2)\times Rate$. \\
		\textit{Step 6}: Circular shift the PDP by $S$. \newline \hspace*{9mm}Conduct measurements based on Algorithm 1\\
		\hline
	\end{tabular}%
	\label{tab:alg2}%
	\vspace{-1 em}
\end{table}%

\section{Comparison of Drift Correction Techniques}\label{sec2}
The methods in this work involving PTP-based synchronization and automated periodic PDP drift correction is compared to other works \cite{Kanhere2023icc,Mac2017sac,novotny2016pamta,ju2022nyu,Fatih2020ojap} in Table \ref{tab:comp}. The correction technique presented here, clearly, is advantageous with minimal hardware addition, while providing accurate correction of the \textit{time drift}. The method eliminates any tedious post-processing and does not require modeling before measurements, thus, speeding up the radio channel modeling process. While the drift correction and PDP alignment may not be exact, such as with Ray-tracing, the error is minimal with PTP and the proposed algorithms.  

\renewcommand{\arraystretch}{1.2}
\begin{table}[htbp]
	\centering
	\caption{Comparison of time drift correction techniques}
	\vspace{-1 em}
	\begin{tabular}{|h|g|f|B|g|e|}
		\hline
		\textbf{\#} & \textbf{Correction method} &\textbf{Applied} & \textbf{Requisites} & \textbf{Possible Errors} & \textbf{Ref} \\
		\hline
		1     & Ray Tracing & Post & Accurate geometric model of environment & Modeling inaccuracy & \cite{Kanhere2023icc} \\
		\hline
		2     & Time-drift Tracking & During \&\newline{} Post & Recorded time drift files \newline measurement-log & Varying Rb clock drift rates & \cite{ju2022nyu} \\
		\hline
		3     & Periodic re-sync. & During & Reconnect Rb clocks mid-measurement & Insufficient time for re-sync. & \cite{novotny2016pamta,Mac2017sac} \\
		\hline
		4     & CCD* and CDEDAR** & Pre \&\newline Post & Prior modeling of clock drift; \newline Channel sounder physical model & Modeling inaccuracy & \cite{Fatih2020ojap} \\
		\hline
		5     & PTP and automated correction & During & None  & PTP sync. inaccuracy & This work \\
		\hline
		\multicolumn{6}{c}{*CCD: correction of clock drift}\\
		\multicolumn{6}{c}{**CDEDAR: correction of delay error due to antenna rotation}
	\end{tabular}%
	\label{tab:comp}%
	\vspace{-1 em}
\end{table}%
\renewcommand{\arraystretch}{1}

\section*{Conclusion}
The paper presented an automatic method to eliminate the \textit{time drift} observed in untethered sliding correlation channel sounders. The novel method leveraged two RPi CM4s to implement PTP over Wi-Fi and synchronize the untethered Rb clocks at the TX and RX to within few microseconds.
The drift in absolute timing, which is the time-dilated PDP circular shift in a sliding correlation channel sounder, was vastly improved using the proposed method (150 samples/hour), compared to several thousand samples/hour without PTP. Next, periodic drift correction eliminated the PDP circular shift between measurement sweeps and achieved absolute timing to sub-nanosecond accuracy. The PTP implementation in various configurations revealed minimizing the network hops was key in achieving greater synchronization accuracy over PTP (Table \ref{tab:ptp}). The presented solution was implemented with minimal additional hardware and can obviate the tedious use of ray-tracing to synthesize omnidirectional PDPs from directional measurements. 

\section*{Acknowledgment}
The authors acknowledge Er. Ashutosh Shrivastava and Er. Prasant Adhikari for their insightful discussions on PTP, Linux, and the network stack.

\bibliographystyle{IEEEtran}
\bibliography{references}

\end{document}